\documentclass[oupdraft]{bio}
\usepackage[colorlinks=true, urlcolor=citecolor, linkcolor=citecolor, citecolor=citecolor]{hyperref}
\usepackage{amsmath, graphicx, amsfonts, mathrsfs, float, enumerate, subcaption, amssymb, gensymb}
\usepackage[utf8]{inputenc}

\pdfoutput=1

\begin{document}

\title{PCA leverage: outlier detection for high-dimensional functional magnetic resonance imaging data}
\author{AMANDA F. MEJIA$^{\rm a}$, MARY BETH NEBEL$^{\rm b}$, ANI ELOYAN$^{\rm c}$, BRIAN CAFFO$^{\rm a}$ and MARTIN A. LINDQUIST$^{\rm a\ast}$\\[4pt]
$^{\rm a}$\textit{Department of Biostatistics, Johns Hopkins University, USA}\\ 
$^{\rm b}$\textit{Center for Neurodevelopmental and Imaging Research, Kennedy Krieger Institute, USA}\\
$^{\rm c}$\textit{Department of Biostatistics, Brown University, USA}\\[2pt]
{mlindqui@jhsph.edu}}\ \\


\markboth%
{A. F. Mejia and others}
{PCA leverage: outlier detection for high-dimensional fMRI data}

\maketitle

\footnotetext{To whom correspondence should be addressed.}

\begin{abstract}
{Outlier detection for high-dimensional (HD) data is a popular topic in modern statistical research.  However, one source of HD data that has received relatively little attention is functional magnetic resonance images (fMRI), which consists of hundreds of thousands of measurements sampled at hundreds of time points.  At a time when the availability of fMRI data is rapidly growing---primarily through large, publicly available grassroots datasets---automated quality control and outlier detection methods are greatly needed.  We propose PCA leverage and demonstrate how it can be used to identify outlying time points in an fMRI run.  Furthermore, PCA leverage is a measure of the influence of each observation on the estimation of principal components, which are often of interest in fMRI data.  We also propose an alternative measure, PCA robust distance, which is less sensitive to outliers and has controllable statistical properties.  The proposed methods are validated through simulation studies and are shown to be highly accurate.  We also conduct a reliability study using resting-state fMRI data from the Autism Brain Imaging Data Exchange and find that removal of outliers using the proposed methods results in more reliable estimation of subject-level resting-state networks using ICA.}{High-dimensional statistics; fMRI; Image analysis; Leverage; Outlier detection; Principal component analysis; Robust statistics}

\end{abstract}

\section{Introduction}

Outliers in high-dimensional (HD) settings, such as genetics, medical imaging and chemometrics, are a common problem in modern statistics and have been the focus of much recent research \citep{hubert2005robpca, filzmoser2008outlier, hadi2009detection, shieh2009detecting, fritsch2012detecting, Ro07062015}.  One such source of HD data is functional magnetic resonance imaging (fMRI). An fMRI run usually contains 100,000-200,000 volumetric elements or “voxels” within the brain, which are sampled at hundreds of time points.  Here, we consider voxels to be variables and time points to be observations, in which case the outlier problem is to identify time points that contain high levels of noise or artifacts.  

Multiple noise sources related to the hardware and the participant \citep{lindquist2008statistical} can corrupt fMRI data, including magnetic field instabilities, head movement, and physiological effects, such as heartbeat and respiration. Noise sources appear as high-frequency “spikes”, image artifacts distortions, and signal drift.  fMRI data also undergo a series of complex preprocessing steps before being analyzed; errors during any one of these steps could introduce additional artifacts.  Thus, performing adequate quality control prior to statistical analysis is critical.

In recent years, the availability of fMRI data has increased rapidly. The emergence of a number of publicly available fMRI databases, often focusing on a specific disease or disorder, presents a great opportunity to study brain function and organization. However, these datasets are usually collected from multiple sites with varying methods for acquisition, processing and quality control, resulting in widely varying levels of quality and high rates of artifacts. In the absence of automated outlier detection methods appropriate for fMRI data, quality inspection often takes place in a manual or semi-automated manner by individual research groups. This presents a timely opportunity for statisticians to develop more automated methods.

Here we propose an HD outlier detection method based on dimension reduction through principal components analysis (PCA) and established measures of outlyingness, namely leverage and robust distances.  While leverage has not typically been employed for outlier identification outside of a regression framework, we argue for leverage as a meaningful measure when the principal components (PCs) are themselves of interest, which is often true for fMRI data.  

Several outlier detection methods for standard and HD data use PCA, including PCA influence functions and other PC sensitivity measures \citep{brooks1994diagnostics, gao2005new}. However, these methods are often computationally demanding as they rely on re-estimating the PCs with each observation left out. Similarly, methods that depend on robust covariance estimation (see \cite{hadi2009detection} for a review) are usually not suited for HD settings.  One such method, the minimum covariance determinant (MCD) estimator, identifies the observation subset with the smallest sample covariance matrix determinant \citep{rousseeuw1985multivariate}. \cite{hubert2005robpca} proposed ROBPCA, a robust PCA method for HD data that can also identify outliers, which lie far from the robust PCs space.  \cite{filzmoser2008outlier} proposed PCOut and Sign, two computationally efficient methods that perform standard PCA after robustly scaling the data and looking for outliers within the principal directions explaining 99\% of the variance.  \cite{Ro07062015} proposed the minimum diagonal product estimator, which is related to the MCD but ignores off-diagonal elements and is identifiable when there are more variables than observations.  \cite{fritsch2012detecting} proposed an HD adaptation of the MCD through regularization and applied the method to neuroimaging summary statistics.

However, such methods are often validated using only moderately sized data containing more observations than variables. One exception comes from \cite{shieh2009detecting} who proposed identifying outlying genes in microarray data by performing PCA dimension reduction prior to robust distance computation on the reduced data. The method was validated on a dataset of approximately 100 observations and 2000 variables and shown to result in fewer false positives and false negatives than ROBPCA.

Existing methods for fMRI artifact identification have focused on head motion and ad-hoc measures of quality. While the removal of affected time points (“scrubbing” or “spike regression”) using these methods appears beneficial \citep{satterthwaite2013improved, power2014methods}, a more unified outlier detection framework is needed, as motion is only one potential artifact source in fMRI data. In addition, existing methods result in a collection of measures that must somehow be combined. We propose a single measure of outlyingness related to the influence of each time point on PC estimation, which is the basis of several common brain connectivity measures (see Section \ref{leverage_def}).

The remainder of this paper is organized as follows.  We begin with a description of our statistical methodology.  We then present a simulation study, which is used to assess the sensitivity and specificity of the proposed methods.  Next, we present a reliability analysis employing the Autism Brain Imaging Data Exchange (ABIDE) dataset.  We conclude with a brief discussion.

\section{Methods}

As described in detail below, we propose two PCA-based measures of outlyingness, \textit{PCA leverage} and \textit{PCA robust distance}, and develop thresholding rules to label outliers using either measure.  For both measures, we begin with PCA dimension reduction.  All computations are performed in the R statistical environment version 3.1.1 \citep{R2014}.

\subsection{Dimension Reduction}
\label{sec:dimred}

Let $T$ be the number of 3-dimensional ``volumes'' collected over time in an fMRI run, and let $V$ be the number of voxels in the brain. Let $\mathbf{Y}_{T\times V}$ ($T\ll V$) represent an fMRI run, where each row of $\mathbf{Y}$ is a vectorized volume.  We first center and scale each column of $\mathbf{Y}$ relative to its median and median absolute deviation \citep{hampel1986robust}, respectively, to avoid the influence of outliers. The singular value decomposition (SVD) \citep{golub1970singular} of $\mathbf{Y}$ is given by $\mathbf{Y}=\mathbf{U}_{T\times T}\mathbf{D}_{T\times T}\mathbf{V}_{T\times V}^t$, where $\mathbf{D}$ is diagonal with elements $d_1 \geq d_2 \geq \cdots \geq d_T \geq 0$ and $\mathbf{U}\mathbf{U}^t=\mathbf{U}^t\mathbf{U}=\mathbf{V}^t\mathbf{V}=\mathbf{I}_T$. Here $\mathbf{A}^t$ denotes the transpose of matrix $\mathbf{A}$. The rows of $\mathbf{V}^t$ contain the PCs or \textit{eigenimages} of $\mathbf{Y}$, and the columns of $\mathbf{\tilde{U}}=\mathbf{UD}$ contain the corresponding PC scores.  Note that to avoid memory limitations, rather than compute the SVD of $\mathbf{Y}$ directly, one generally computes the SVD of $\mathbf{Y}\mathbf{Y}^t$ to obtain $\mathbf{U}$ and $\mathbf{D}$ and then solves for $\mathbf{V}^t$.  

We retain $Q<T$ principal components, so that the ``reduced data'' are given by the submatrices of $\mathbf{U}$ and $\mathbf{D}$ corresponding to the first $Q$ principal components.  For ease of notation, we redefine $\mathbf{U}_{T\times Q}$ and $\mathbf{D}_{Q\times Q}$ to represent these submatrices and $\mathbf{\tilde{U}}_{T\times Q}=\mathbf{UD}$.  To choose the model order $Q$, we retain only components with a greater-than-average eigenvalue. While more sophisticated cutoff methods exist, we find that this simple cutoff rule works well in practice \citep{jackson1993stopping}. To avoid extreme solutions, we require $15\leq Q\leq50$.


\subsection{Principal components analysis leverage}
\label{leverage_def}

In regression, leverage is defined as the diagonals of the ``hat matrix'' $\mathbf{H}=\mathbf{X}(\mathbf{X}^t\mathbf{X})^{-1}\mathbf{X}^t$, where $\mathbf{X}$ is a matrix of explanatory variables \citep{neter1996applied}.  The hat matrix projects the outcome variable(s) $\mathbf{Y}$ onto the column space of $\mathbf{X}$, yielding the projected data $\mathbf{\hat{Y}}=\mathbf{HY}$.  Leverage, bounded between $0$ and $1$, is often used to assess the potential influence of an observation on the regression fit, as it is the change in $\hat{y}_i$ due to a 1-unit change in $y_i$ and is proportional to the uncertainty in the $\hat{y}_i$ estimate, since $\text{Var}(\hat{\mathbf{Y}})=\sigma^2\mathbf{H}$.  Particularly relevant to our context, leverage is also a measure of outlyingness among the explanatory variables, as it is related to the Mahalanobis distance.

\begin{figure}[H]
\centering
\includegraphics[width=5in, page=1, trim=0 10mm 5mm 0, clip]{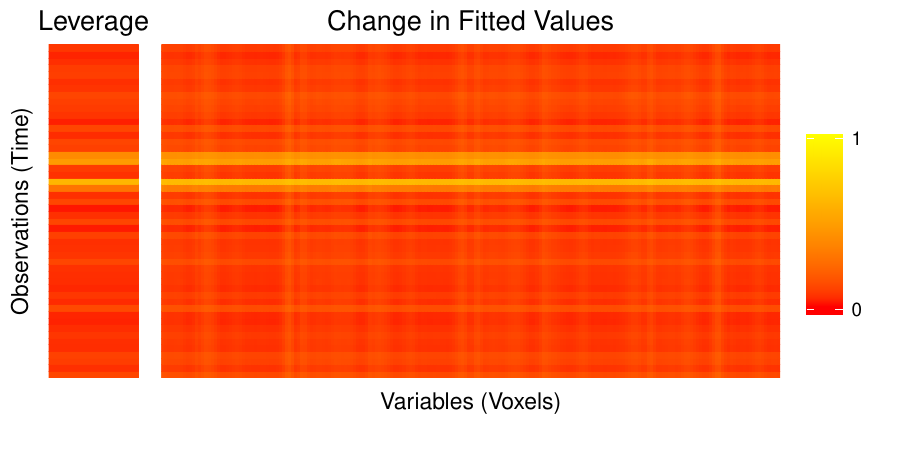}
\includegraphics[width=4.5in]{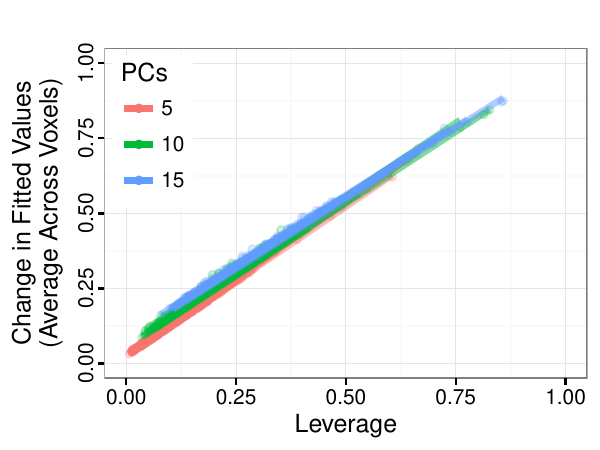}
\caption{\textbf{Top panel. }For one randomly sampled subject, 50 contiguous time points and 200 contiguous voxels were randomly selected.  For the resulting dataset $\mathbf{Y}_{50\times200}$, five PCs were identified, and PCA leverage was computed for each time point $t$, displayed in the column on the left.  After centering and scaling $\mathbf{Y}$, each observed value ${Y}_{[t,v]}$ was increased by one unit and PCs and scores recomputed.  The matrix displayed on the right shows the resulting \textit{change} in the fitted value $\hat{Y}_{[t,v]}$, where $\hat{\mathbf{Y}}=\mathbf{UDV}^t$. Although some variation is seen across voxels, the observed change in fitted values is overall quite similar to the leverage.  Other randomly sampled subjects show similar patterns, supporting the analogy with regression (where the relationship is exact) and the concept of PCA leverage as a measure of influence in PCA.  \textbf{Bottom panel. } We performed the analysis described above for $100$ randomly sampled subjects, then computed the average change in fitted values across voxels.  We performed the analysis with $5$, $10$ and $15$ PCs retained.  The plot displays the PCA leverage and average change in fitted values for each subject, as well as a linear smoother across subjects.  PCA leverage and the average change in fitted values are nearly equal, again supporting PCA leverage as a measure of influence in PCA.}
\label{change_Yhat}
\end{figure}

Extending leverage to the PCA context, we treat $\mathbf{\tilde{U}}=\mathbf{UD}$ as an estimated design matrix in the estimation of $\mathbf{V}^t$.  With $\mathbf{U}$ and $\mathbf{D}$ fixed, $\mathbf{V}^t=\mathbf{D}^{-1}\mathbf{U}^t\mathbf{Y}$ is equivalent to the least squares estimate $\mathbf{\hat{V}}^t$ in the multivariate regression model $\mathbf{Y}=\mathbf{\tilde U}\mathbf{V}^t+\mathbf{E}$.  We therefore define PCA leverage as $\mathbf{h}=\{h_1,\dots,h_T\}=\text{diag}\{\mathbf{H}\}$, where $\mathbf{H}=\mathbf{\tilde{U}}(\mathbf{\tilde{U}}^t\mathbf{\tilde{U}})^{-1}\mathbf{\tilde{U}}^t = \mathbf{UU}^t$.  Note that $\mathbf{D}$ is simply a scaling factor applied to each variable and therefore has no effect on leverage.  Continuing the regression analogy, in PCA, $\mathbf{H}$ projects $\mathbf{Y}$ onto the column space of $\mathbf{\tilde{U}}$, the principal directions, as $\mathbf{\hat{Y}}=\mathbf{UDV}^t=\mathbf{UDD}^{-1}\mathbf{U}^t\mathbf{Y}=\mathbf{UU}^t\mathbf{Y}=\mathbf{HY}$.  Furthermore, PCA leverage is a measure of outlyingness among the PCA scores and within $\mathbf{\hat{Y}}=\mathbf{UDV}^t$, since $\mathbf{\hat{Y}}(\mathbf{\hat{Y}}^t\mathbf{\hat{Y}})^{-1}\mathbf{\hat{Y}}^t = \mathbf{UU}^t=\mathbf{H}$. While in reality $\mathbf{U}$ and $\mathbf{D}$ are not fixed and PCA leverage therefore only approximately represents the influence of each observation on the PCs and fitted values in $\mathbf{\hat{Y}}$, we find this approximation to be quite close in practice, as illustrated in Figure \ref{change_Yhat}.  Note that dimension reduction is essential for PCA leverage to be informative, since $\mathbf{U}\mathbf{U}^t=\mathbf{I}$ when all $T$ components are retained.   

In regression, leverage only represents the \textit{potential} influence of an observation on regression coefficient estimation; influence points must be outliers in the explanatory variables (``leverage points'') as well as in the response variable(s).  In contrast, PCA leverage is a more direct measure of influence, as PCA leverage points are outliers in both $\mathbf{\tilde U}$ and the original data $\mathbf{Y}$. Thus, PCA leverage points are also influence points.  Furthermore, while in regression we discern ``good'' from ``bad'' leverage points, fMRI observations with high PCA leverage are unlikely to represent true signal, since the signal change associated with neuronal sources is very small compared with noise and artifacts.  Therefore, we assume that all observations with high PCA leverage are ``bad'' influence points in the fMRI context.  

Moreover, the interpretation of PCA leverage as the influence of each observation on PC estimation is particularly relevant for resting-state fMRI (rs-fMRI). PC estimation is a preprocessing step for one of the most common types of analysis for rs-fMRI data: estimation of spatially independent brain networks and the functional connectivity between those networks.  In such analyses, PCA leverage is both a measure of influence on the quantity of interest and of outlyingness.

In setting a leverage threshold to identify outliers, it is important to recognize that the sum of leverages for a set of observations equals the number of variables in the design matrix.  The mean leverage of all $T$ observations is therefore fixed at $Q/T$.  Outliers wield a large amount of leverage, and their presence reduces the leverage of all remaining observations, such that the mean $Q/T$ may be significantly greater than the leverage of normal observations.  Thus, the median is a more appropriate reference quantity than the mean for normal observations.  If the leverage of observation $t$ exceeds $\alpha$ times the median $m_h=\text{med}(h_1,\dots,h_T)$, it is labeled a ``leverage outlier''.  In the simulations and experimental data analysis described below, we consider $\alpha\in[3,8]$.

While such practical rules may work well in the absence of convenient statistical properties, a formal statistical test for outliers with known and controllable properties is desirable. In the following section, we propose an alternative robust distance measure based on minimum covariance determinant (MCD) estimators \citep{rousseeuw1985multivariate}. 

\subsection{Principal components robust distance}

For a design matrix with an intercept or centered variables, leverage is related to the squared empirical Mahalanobis distance \citep{mahalanobis1936generalized}, defined for an $n\times p$ matrix $\mathbf{X}$ and observation $i$ as $d_i^2=(X_i-\bar{X})^tS^{-1}(X_i-\bar{X})$, where $\bar{X}$ and $S$ are the sample mean and covariance matrix of $\mathbf{X}$, respectively.  The Mahalanobis distance is known to be sensitive to outliers due to their influence on the sample mean and covariance, which may lead to ``masking'', a phenomenon in which truly outlying observations appear normal due to the presence of more extreme outliers \citep{rousseeuw1990unmasking,rousseeuw2011robust}.


As an alternate measure, we adopt the minimum covariance determinant (MCD) distance proposed by \cite{rousseeuw1985multivariate}.  For a general dataset, let $n$ be the number of observations and $p$ be the number of variables.  The MCD estimators of location, $\bar{X}^*$, and scale, $S^*$, are obtained by computing the sample mean and covariance within a subset of the data of size $h<n$ for which the confidence ellipsoid determined by $S^*$ and centered at $\bar{X}^*$ has minimal volume.  The maximum breakdown point of MCD estimators is obtained by setting $h=\lfloor (n+p+1)/2\rfloor$ and approaches $50\%$ as $n\to\infty$.  The MCD distance $d_{S^*}^2(X_i,\bar{X}^*)$ is then computed as a Mahalanobis distance using $\bar{X}^*$ and $S^*$ in place of $\bar{X}$ and $S$.  For ease of notation, let $d_i^2\equiv d_{S^*}^2(X_i,\bar{X}^*)$.

Let $\mathcal{N}={1,\dots,n}$, and let $\mathcal{N}^*$, $|\mathcal{N}^*|=h$, be the indices of the observations selected to compute $\bar{X}^*$ and $S^*$.  Let $\mathcal{N}^-=\mathcal{N}\setminus\mathcal{N}^*$ be the indices of the remaining observations, among which we look for outliers.  For Gaussian data, $\{d_i^2:i\in\mathcal{N}^*\}$ approximately follow a $\chi^2_p$ distribution \citep{hubert2005robpca, shieh2009detecting}, while for $i\in\mathcal{N}^-$,
$$
\tilde{d}_i^2 := \frac{c(m-p+1)}{pm} d_i^2 \sim F_{p,m-p+1},
$$
where $c$ and $m$ can be estimated asymptotically or through simulation. (While some previous work has simply assumed a $\chi^2_p$ distribution for $i\in\mathcal{N}$, we find this to result in many false positives.) To estimate $c$ we use the asymptotic form, $\hat{c}=Pr\left\{\chi_{p+2}^2<\chi_{p,h/n}^2\right\}\big/(h/n)$, while to estimate $m$ we use the small sample-corrected asymptotic form given in \cite{hardin2005distribution}.  To improve the F-distribution fit, \cite{maronna2002robust} and \cite{filzmoser2008outlier} scale the distances to match the median of the theoretical distribution. However, as $\mathcal{N}^-$ contains at most half of the original observations, the median within $\mathcal{N}^-$ represents the $75$th or greater quantile within $\mathcal{N}$ and may be contaminated with outliers.  Therefore, we let $\tilde{\tilde{d}}_i^2 := \tilde{d}_i^2 {F_{p,m-p+1,0.1}}/q_{0.1}$, where $q_{0.1}$ is the $10$th sample quantile of $\{\tilde{d}_i^2:i\in\mathcal{N}^-\}$. We label a ``distance outlier'' any observation in $\mathcal{N}^-$ with $\tilde{\tilde{d}}_i^2$ greater than the $(1-\gamma)$th quantile of the theoretical F distribution.  In our simulations and experimental data analysis, we consider $1-\gamma$ ranging from $99$ to $99.99$.

For time series data, where the assumption of independence among observations is violated, the distributional results given in \cite{hardin2005distribution} may be invalid.  As the autocorrelation in fMRI time series is often modeled as an AR(1) process with a coefficient of $0.3$, we divide each fMRI time series into three subsets, each consisting of every third observation.  The autocorrelation within each subset is negligible at approximately $0.3^3 = 0.027$.  To obtain the MCD distance for each observation, we use the MCD estimates of center and scale within each subset, averaged across subsets, and we find that this significantly improves the distributional fit.

\section{Simulation Study}

\subsection{Construction of baseline scans}

Our simulated dataset is based on fMRI scans from three subjects collected as part of the ABIDE dataset (described in Section \ref{data_section}).  For generalizability, each subject was chosen from a different data collection site.  For each scan, we identify a contiguous subset of volumes containing no detectable artifacts, resulting in 141, 171 and 89 volumes, respectively.  We reduce dimensionality by using only the 45th axial (horizontal) slice, corresponding roughly to the center of the brain.  For scan $i=1,2,3$, let $T_i$ be the resulting length of the scan and $V_i$ be the resulting number of voxels in the brain mask, so that scan $i$ is represented by the $T_i\times V_i$ matrix $\mathbf{Y}_i$.  We can separate $\mathbf{Y}_i$ into an anatomical baseline, the mean image $\mathbf{b}_i$ ($V_i\times1$), and the residual $\mathbf{Z}_i$, representing primarily functional information.  Then $\mathbf{Y}_i = \mathbf{Z}_i + \mathbf{1}\mathbf{b}_i^t$, where $\mathbf{1}$ is a vector of $1$s of length $T_i$.  

We then use independent components analysis (ICA), a blind-source separation algorithm, to decompose the intrinsic activity in $\mathbf{Z}_i$ into a number of spatially independent sources (\cite{mckeown1997analysis}) (described in Section \ref{data_section}).  Let $Q_i$ be the number of sources of neuronal signal identified for scan $i$.  Then $\mathbf{Z}_i = \mathbf{A}_i\mathbf{S}_i + \mathbf{E}_i$, where $\mathbf{S}_i$ ($Q_i\times V_i$) contains the spatial maps of each source, and $\mathbf{A}_i$ ($T_i\times Q_i$) contains the time courses of each source.  The residual $\mathbf{E}_i$ contains structured (spatially and temporally correlated) noise.  Let $\mathbf{X}_i=\mathbf{A}_i\mathbf{S}_i$.
 
\subsection{Artifact-free images}

For each scan $i$, we construct three simulation setups: baseline image ($\mathbf{B}_i=\mathbf{1}\mathbf{b}_i^t$) plus white noise (setup 1); baseline image plus functional signal ($\mathbf{B}_i + \mathbf{X}_i$) plus white noise (setup 2); and baseline image plus functional signal plus structured noise ($\mathbf{B}_i + \mathbf{X}_i + \alpha\mathbf{E}_i$) (setup 3).

To test the specificity of each outlier detection method in the artifact-free setting, we generate images with varying signal-to-noise ratio (SNR) in the following way. For scan $i$, we have true signal variance $\hat\sigma^2_{i,X}=\frac{1}{V_i}\sum_{v=1}^{V_i} \hat{Var}\{X_i(v)\}$ and true noise variance $\hat\sigma^2_{i,E}=\frac{1}{V_i}\sum_{v=1}^{V_i} \hat{Var}\{E_i(v)\}$.  Defining SNR as the ratio of signal variance to noise variance, let $\lambda\in\{$0.025, 0.050, 0.075, 0.1, 0.2, 0.4, 0.6, 0.8, 1.0$\}$ be the desired SNR of the simulated scans.  For setups 1 and 2, we generate the white noise matrix $\mathbf{W}_i(\lambda)$ for scan $i$ as independent, mean-zero Gaussian noise with variance $\sigma^2_{i,E}(\lambda) = \hat\sigma^2_{i,X}/\lambda$.  For setup 3, we generate the structured noise matrix $\mathbf{E}_i(\lambda)=\sqrt{SNR_i/\lambda}\times\mathbf{E}_i$, where $SNR_i=\hat\sigma^2_{i,X}/\hat\sigma^2_{i,E}$ is the baseline SNR of scan $i=1,2,3$, equal to $0.063$, $0.050$ and $0.048$, respectively.  Therefore, the simulated artifact-free data at SNR $\lambda$ is $\mathbf{B}_i + \mathbf{W}_i(\lambda)$ for setup 1; $\mathbf{B}_i + \mathbf{X}_i + \mathbf{W}_i(\lambda)$ for setup 2; and $\mathbf{B}_i + \mathbf{X}_i + \mathbf{E}_i(\lambda)$ for setup 3.  For setups 1 and 2, we randomly generate $\mathbf{W}_i(\lambda)$ $1000$ times; for setup 3, the noise is fixed.

The specificity, or percentage of observations \textit{not} labeled as outliers that are truly non-outliers, in this case is simply the percentage of volumes in each scan not labeled as outliers.  Figure \ref{spec_noart} shows the mean specificity across $1000$ iterations, where each line represents a scan and SNR level.  The lines in red correspond to SNR of $0.05$, which is close to the observed SNR of each scan.  Specificity is nearly $100\%$ for both leverage and robust distance methods across all thresholds and SNR levels considered.  In the presence of structured noise, the specificity of the robust distance method is approximately $97$-$98\%$ in some cases, unless the $99.99$th quantile threshold is used.

\subsection{Images with artifacts}
\label{sec:artifacts}

We generate four common types of fMRI artifacts: spikes, motion, banding, and ghosting.  Spikes are created by increasing the intensity of an entire volume by a given percentage.  Motion artifacts are created by rotating a volume by a given angle.  Banding artifacts are generated by changing an intensity in the Fourier transform of the image, resulting in a striped appearance.  Ghosting artifacts are created by superimposing a figure or ``ghost'' moving through space over time.  

\begin{figure}[H]
\centering
\begin{subfigure}[b]{1\textwidth}
\centering\includegraphics[scale=0.36, page=1, trim=0 0 0 5mm, clip]{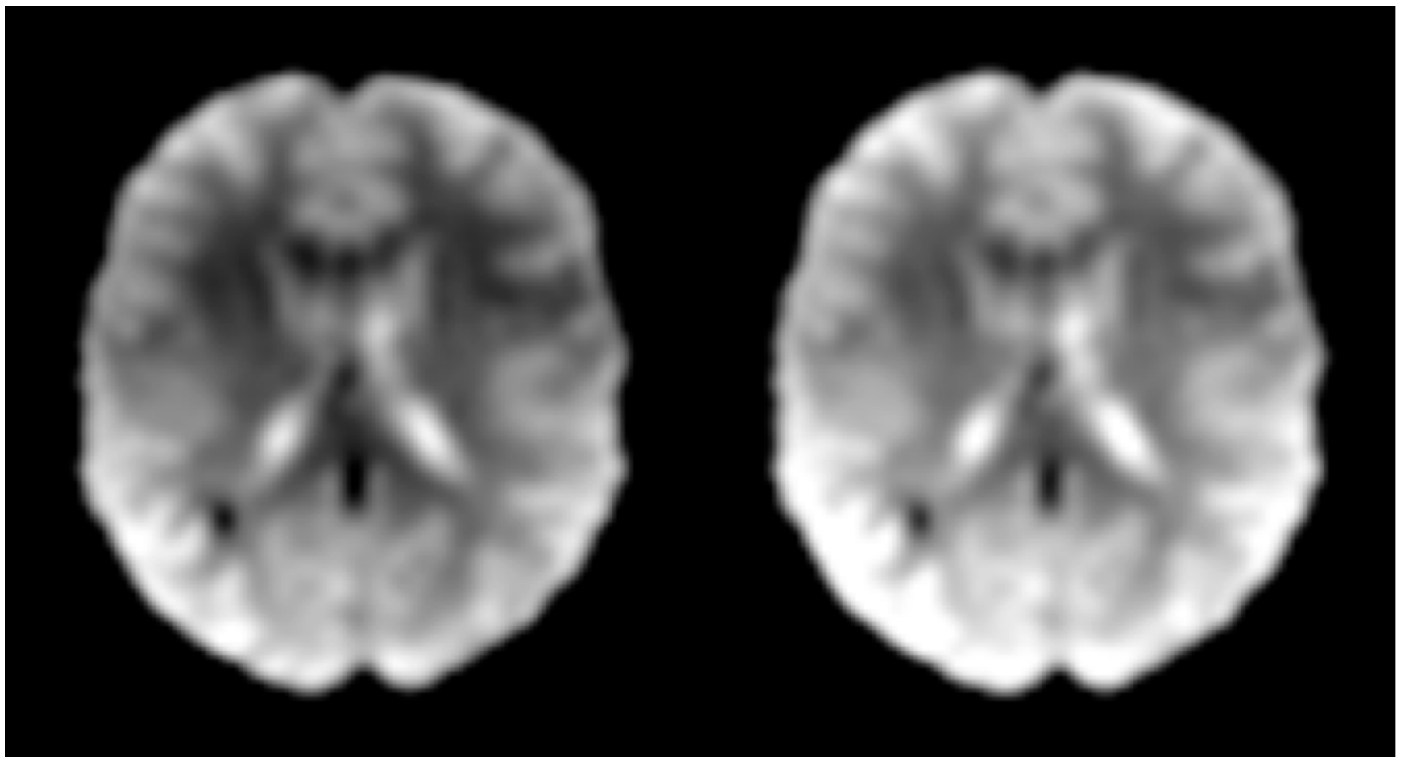}
\caption{Spike artifact\\[10pt]}
\end{subfigure}
\begin{subfigure}[b]{0.3\textwidth}
\centering\includegraphics[scale=0.36, page=2]{images.pdf}
\caption{Rotation artifact}
\end{subfigure}
\begin{subfigure}[b]{0.3\textwidth}
\centering\includegraphics[scale=0.36, page=3]{images.pdf}
\caption{Banding artifact}
\end{subfigure}
\begin{subfigure}[b]{0.3\textwidth}
\centering\includegraphics[scale=0.36, page=4]{images.pdf}
\caption{Ghosting artifact}
\end{subfigure}
\caption{Examples of each artifact type.  Figure (a) shows a normal volume on the left and a volume with a spike artifact on the right.  Figure (b) shows the image mask before and after rotation.  The spike, rotation and ghosting artifacts are generated from the maximum artifact intensity as described in Section \ref{sec:artifacts}; the banding artifact is generated randomly as described in Section \ref{sec:artifacts}.\\[10pt]}
\label{artifacts}
\end{figure}

\begin{figure}[H]
\centering
\includegraphics[width=6in, trim=0 0.7in 0 0, clip]{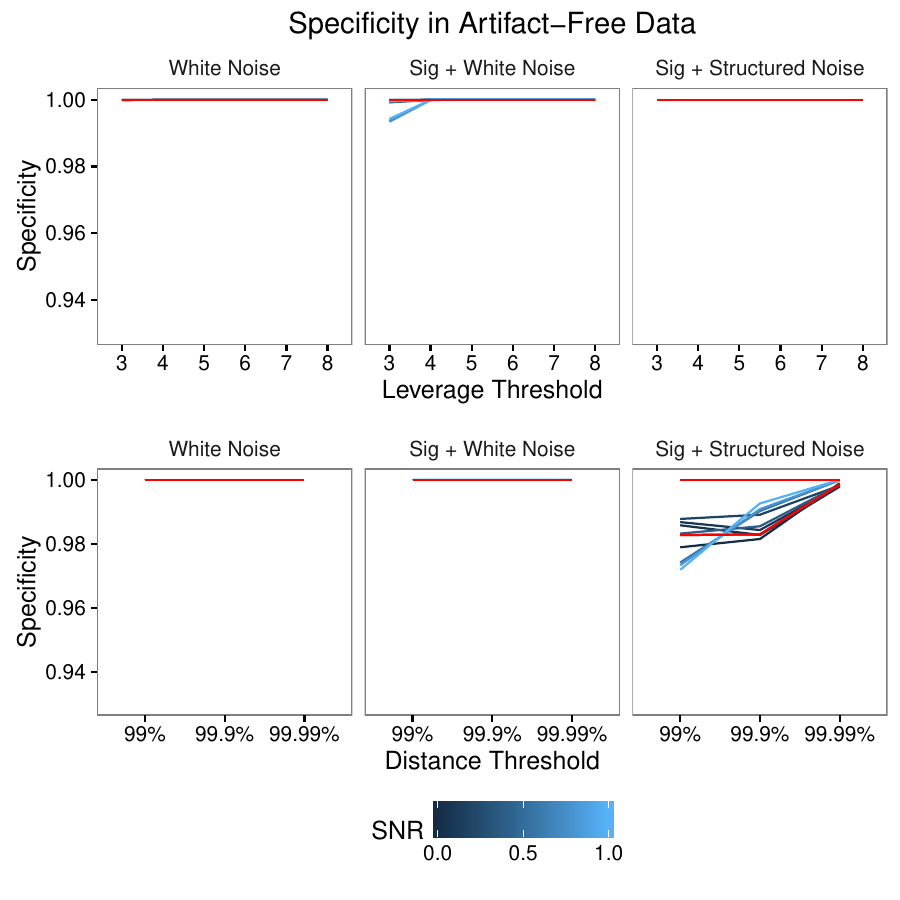}\\
\includegraphics[scale=0.8]{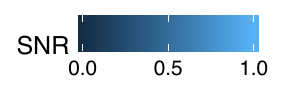}
\caption{Specificity of each method in the \textit{absence} of artifacts by simulation setup.  Each line shows the mean across 1000 iterations for a given scan $i=1,2,3$ and SNR.  The lines in red correspond to SNR of 0.05, which is close to the observed SNR of the fMRI scans used to construct the simulated scans.}
\label{spec_noart}
\end{figure}

At each of $1000$ iterations, one simulated fMRI scan is generated for each subject, SNR level, artifact type and simulation setup.  For spike, motion and banding artifacts, 10 volumes are randomly selected from each scan, and the artifact intensity for each volume is generated from a uniform distribution (range $1\%-10\%$ percent intensity increase for spike artifacts; $1\degree-5\degree$ rotation for motion artifacts; $50-200$ times change at location $(15,15)$ of the Fourier transformed image).  For ghosting artifacts, 9 sequential volumes are randomly selected, and the mean intensity of the ghost, relative to the mean intensity of the image, is randomly generated from a uniform distribution (range $0.06-0.32$). An example of each artifact type is displayed in Figure \ref{artifacts}.

\begin{figure}[H]
\centering
\includegraphics[width=6in, trim=0 0.7in 0 0, clip]{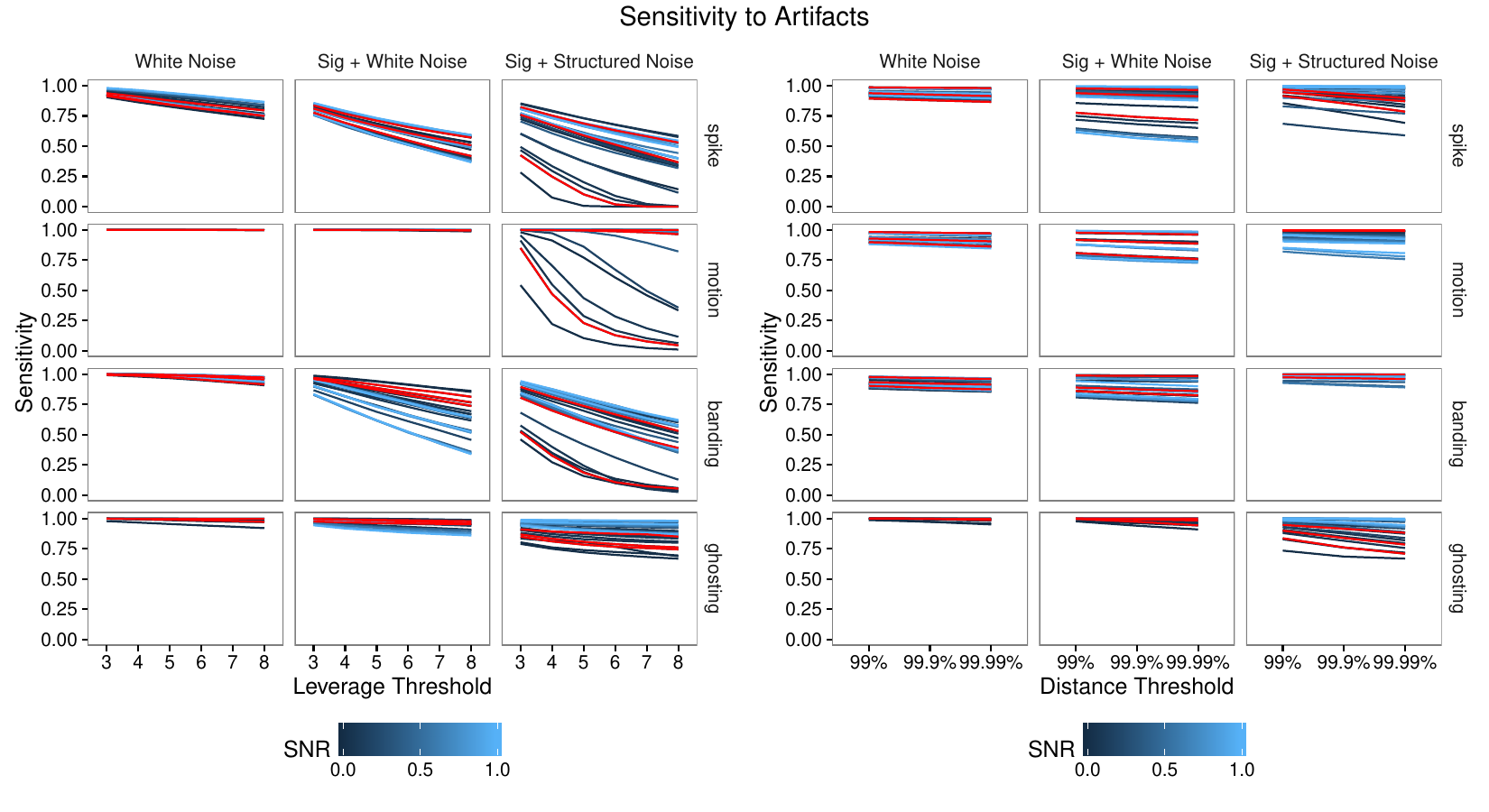}\\[10pt]
\includegraphics[width=6in, trim=0 0.7in 0 0, clip]{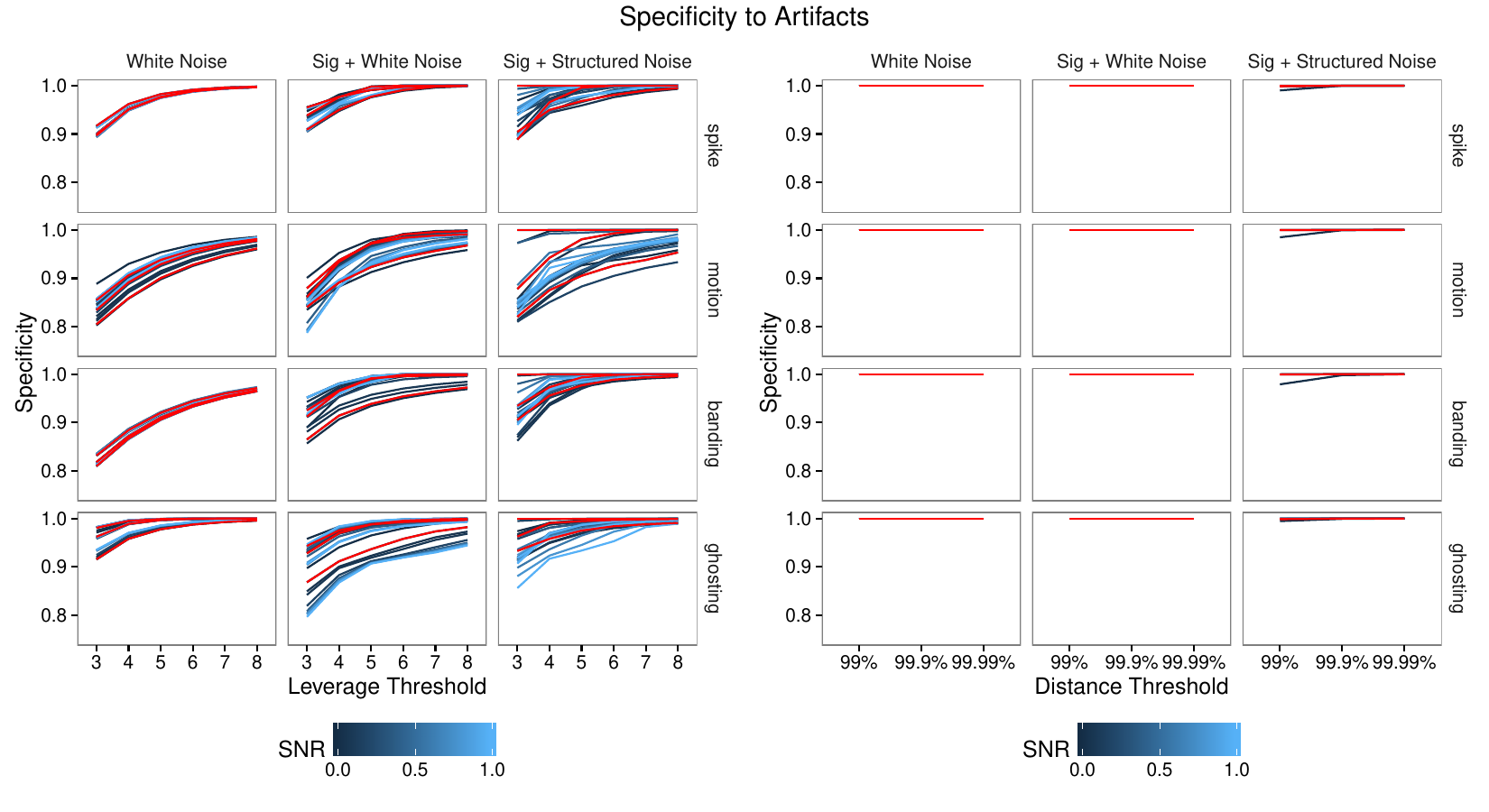}
\includegraphics[scale=0.8]{SNR_legend.pdf}
\caption{Sensitivity and specificity of each method in the \textit{presence} of artifacts by simulation setup.  Each line shows the mean across 1000 iterations for a given scan $i=1,2,3$ and SNR. The lines in red correspond to SNR of 0.05, which is close to the observed SNR of the fMRI scans used to construct the simulated scans.}
\label{specsens}
\end{figure}

We are interested in both the specificity and the sensitivity, or the percentage of true outliers identified as outliers.  Figure \ref{specsens} shows the mean sensitivity and specificity for each outlier detection method, simulation setup, and artifact type, where each line represents a scan and SNR level.  The realistic SNR of $0.05$ is shown in red.  As the simulation setup becomes more realistic, the sensitivity to outliers tends to decrease, while the specificity is relatively stable.  The robust distance method has nearly $100\%$ specificity in all scenarios and tends to display higher sensitivity than the leverage method, particularly for banding and spike artifacts.  While differences across artifact types are apparent, these are likely driven by the range of intensities chosen.

\section{Experimental Data Results}
\label{data_section}

Using a large, multi-site fMRI dataset, we assess the result of outlier removal on the scan-rescan reliability of a common type of analysis.  This section is organized as follows.  We begin with a description of the dataset employed and show an example.  We then describe the reliability analysis.  Finally, we quantify the improvement to reliability with the proposed outlier detection methods using a linear mixed model to account for subject and site effects.

\subsection{fMRI Dataset}

ABIDE is a publicly available resource of neuroimaging and phenotypic information from 1112 subjects consisting of 20 datasets collected at 16 sites \citep{di2014autism}. Fully anonymized data from 91 children collected at Kennedy Krieger Institute after the ABIDE release were also included. Image acquisition parameters and demographic information are available at \url{http://fcon\_1000.projects.nitrc.org/indi/abide/}.  For each subject, a $T_1$-weighted MPRAGE volume and one or more rs-fMRI sessions were collected on the same day.  Details of data pre-processing and quality control are provided in Supplementary Materials Appendix A, where Table 1 lists the number of subjects in each dataset. 

\begin{figure}[H]
\centering
\includegraphics[width=5.7in, trim=0.3in 0 0.2in 0, clip]{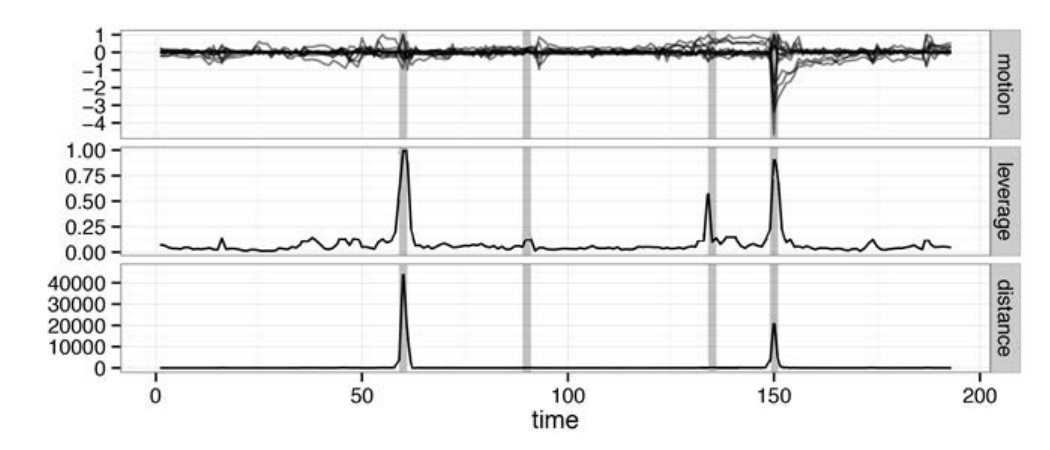}\\
\includegraphics[scale=0.28, trim=0 2.5in 0 0, clip, page=3]{0050048.pdf}
\includegraphics[scale=0.28, trim=0 2.5in 0 0, clip, page=4]{0050048.pdf}\\[10pt]
\includegraphics[scale=0.28, trim=0 2.5in 0 0, clip, page=2]{0050048.pdf}
\includegraphics[scale=0.28, trim=0 2.5in 0 0, clip, page=1]{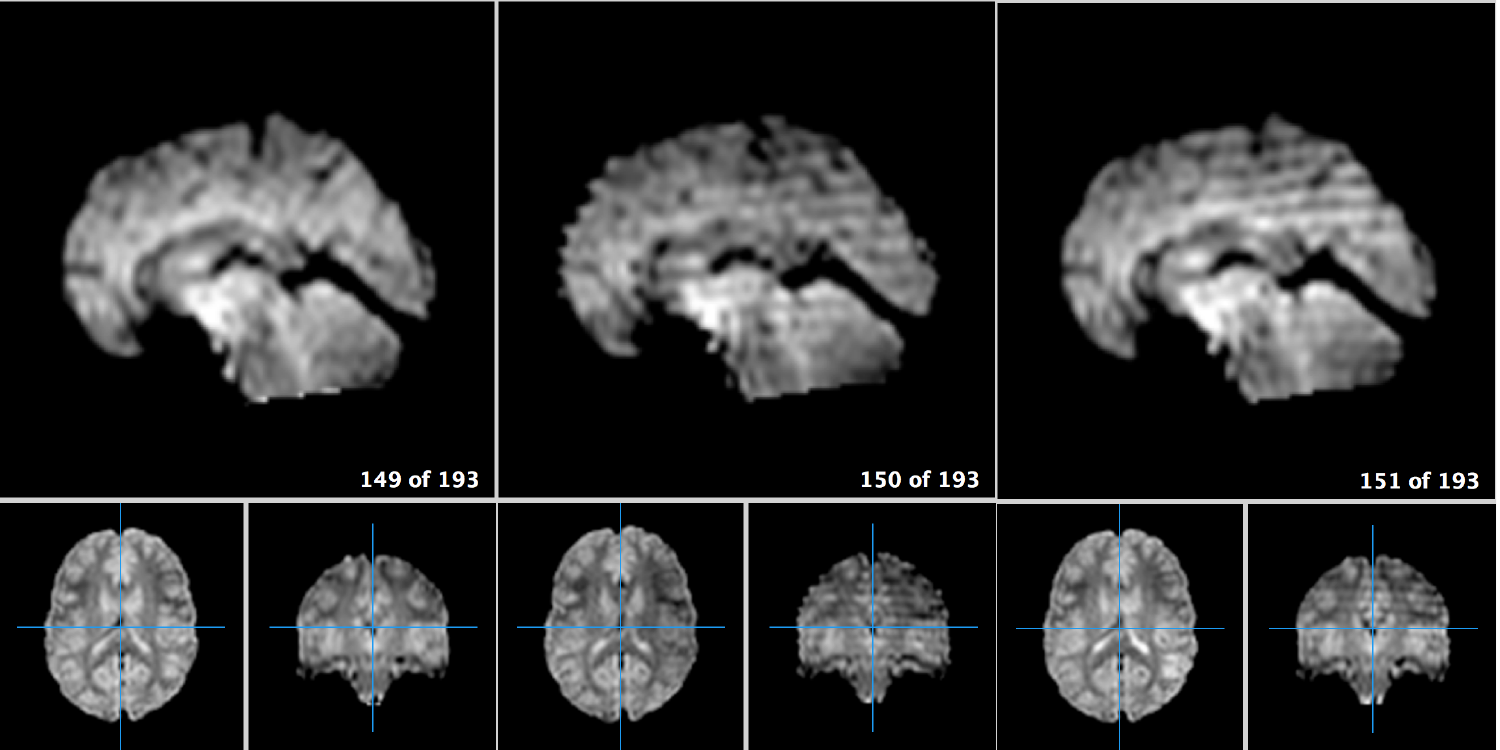}
\caption{For a single subject, the motion parameters, leverage function, and robust distance function.  Below the plot, the volumes corresponding to the spikes at time points 60, 90, 134 and 150 (shaded on the plot) are shown.  Three of the spikes are leverage and distance outliers using any of the thresholds considered ($\alpha\in[3,8]$ for leverage; $1-\gamma\in[99,99.99]$ for robust distance), while the spike at time point 90 is only a leverage outlier at $\alpha=3$.  Obvious banding artifacts are seen at time points 60 and 150, a moderate banding artifact is seen at time point 134, and no visible artifact is apparent at time point 90.  While the artifact at time point 150 would be detected using motion measures, the other spikes would likely go undetected using only motion.}
\label{fig_example}
\end{figure}

For a single example scan, Figure \ref{fig_example} shows the leverage and robust distance functions, along with 6 motion parameters (roll, pitch, yaw, and translation in each direction) and their derivatives, which are commonly used for artifact detection.  Below the plot, the volumes corresponding to the spikes at time points 60, 90, 134 and 150 are shown.  Three of the spikes are leverage and distance outliers using any of the thresholds considered ($\alpha\in[3,8]$ for leverage; $1-\gamma\in[99,99.99]$ for robust distance), while the spike at time point 90 is only a leverage outlier at $\alpha=3$.  Obvious banding artifacts are seen at time points 60 and 150, a moderate banding artifact is seen at time point 134, and no visible artifact is apparent at time point 90.  While the artifact at time point 150 would be detected using motion measures, the other spikes would likely go undetected.


\subsection{Estimation of subject-level brain networks and connectivity}

Resting-state brain networks represent regions of the brain that act in a coordinated manner during rest.  While such networks have traditionally been identified at the group level, there is growing interest in estimating these networks at the subject level, where the higher levels of noise make accurate estimation difficult.  There is also interest in estimating the subject-level ``functional connectivity'', or temporal dependence of neuronal activation, between these different networks \citep{van2010exploring}.  We assess the benefits of outlier removal on the reliability of these networks and their functional connectivity.  Details of the estimation of subject-level resting-state networks are provided in Supplementary Materials Appendix B.  Here we briefly describe the procedure.  We begin by performing group ICA (GICA) separately for each of the ABIDE datasets $k=1,\dots,20$.  The result of GICA is the ${Q\times V_k}$ matrix $\mathbf{S}_k$, where $V_k$ is the number of voxels in the group-level brain mask and $Q=30$ is the number of independent components (ICs).  Each row of $\mathbf{S}_k$ may represent a source of noise (e.g. motion, respiration) or a resting-state network.  After identification of those ICs corresponding to resting-state networks, let $\tilde{\mathbf{S}}_k$ denote the $Q_k\times V_k$ matrix containing the $Q_k$ resting-state networks identified for dataset $k$.  Using dual regression \citep{beckmann2009group}, we obtain $\mathbf{Y}_i \approx \mathbf{A}_i \mathbf{S}_i$, where $\mathbf{S}_i$ is the $Q_k\times V_k$ matrix whose rows contain the estimated resting-state brain networks for subject $i$, and $\mathbf{A}_i$ is the $T_i\times Q_k$ ``mixing matrix'' representing the activation of each network over time.  We are interested in reliable estimation of two quantities: $\mathbf{S}_i$ and the $Q_k\times Q_k$ matrix $Cor(\mathbf{A}_i)$, which represents the functional connectivity between each pair of networks.

\subsection{Measuring reliability of subject-level brain networks and functional connectivity}

Let $\mathbf{S}_{i1}$ and $\mathbf{S}_{i2}$ be two sets of estimated resting-state networks for subject $i$ obtained by performing dual regression separately for two different scanning sessions of subject $i$ in dataset $k$.  There is no need to match components between $\mathbf{S}_{i1}$ and $\mathbf{S}_{i2}$, since the ICs in each correspond to the same group-level ICs in $\tilde{\mathbf{S}}_k$.  To assess reliability, for each subject $i$ and component $q$ we compute the number of overlapping voxels between $\mathbf{S}_{i1}(q)$ and $\mathbf{S}_{i2}(q)$ after both have been thresholded (as described in Supplementary Materials Appendix B).  We then average over all $Q_k$ networks to obtain the average scan-rescan overlap for each subject per network, denoted $Z_{ikm}$ for subject $i$ in dataset $k$ using outlier removal method $m$.  Outlier removal methods include no outlier removal, leverage-based outlier removal with $\alpha=\{3,4,5,6,7,8\}$, and robust distance-based outlier removal with $1-\gamma=\{99,99.9,99.99\}$.  Similarly, to assess reliability of functional connectivity between networks, let $\mathbf{A}_{i1}$ and $\mathbf{A}_{i2}$ be the mixing matrices corresponding to $\mathbf{S}_{i1}$ and $\mathbf{S}_{i2}$.  We compute the mean squared error (MSE) between the upper triangles of $Cor(\mathbf{A}_{i1})$ and $Cor(\mathbf{A}_{i2})$ and denote the result $Y_{ikm}$ for subject $i$ in dataset $k$ using outlier removal method $m$.

Although most subjects in the ABIDE dataset have only a single scanning session, we can simulate scan-rescan data by splitting each subject's data into two contiguous subsets consisting of the first and second half of time points, respectively.  We use the resulting pseudo scan-rescan data to obtain $\mathbf{S}_{i1}$, $\mathbf{S}_{i2}$, $\mathbf{A}_{i1}$ and $\mathbf{A}_{i2}$, then compute our reliability measures $Z_{ikm}$ and $Y_{ikm}$.  While this approach may produce an optimistic estimate of the true scan-rescan reliability, this is not a concern as we are primarily interested in the \textit{change} in reliability due to outlier removal.

To test for changes in reliability of resting-state networks or functional connectivity due to outlier removal, we fit a linear mixed effects model with a fixed effect for each outlier removal method, a fixed effect for each dataset, and a random effect for each subject.  We employ this model for its ability to test several groups and methods simultaneously and to account for within-subject correlation across methods.  Using the $1091$ subjects for whom at least one outlier was identified using any method, we estimate the following model for $M_{ikm}\in\{Y_{ikm},Z_{ikm}\}$:
\begin{align*}
M_{ikm} &= b_{i0} + \gamma_k + \alpha_mI_{m>0} + \epsilon_{ikm},\ 
\epsilon_{ikm}\sim N(0,\sigma^2),\ b_{i0}\sim N(0, \tau^2),
\end{align*}
where $m=0$ indicates no outlier removal.  Here, $\gamma_k$ represents the baseline reliability for subjects in dataset $k$ when no outlier removal is performed, and $\alpha_m$ represents the change in reliability when outlier removal method $m$ is used. To obtain coefficient estimates, we fit this model using the \verb+lme+ function from the \verb+nlme+ package \citep{nlme}.  Since we have a large sample size, we compute Normal 95\% confidence intervals. 

\subsection{The effect of outlier removal on reliability}

\begin{figure}[H]
\centering
\begin{subfigure}[b]{1\textwidth}
\centering\includegraphics[width=6in]{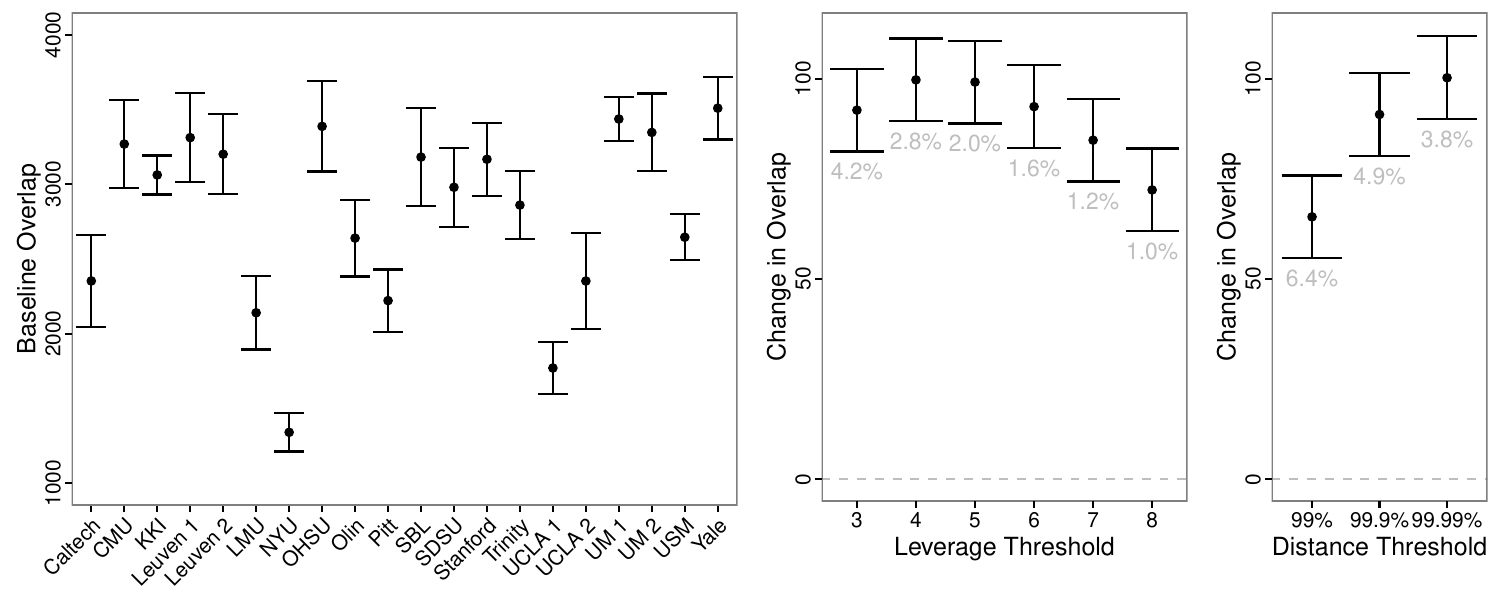}
\caption{Reliability of Brain Networks\\[10pt]}
\end{subfigure}
\begin{subfigure}[b]{1\textwidth}
\centering\includegraphics[width=6in]{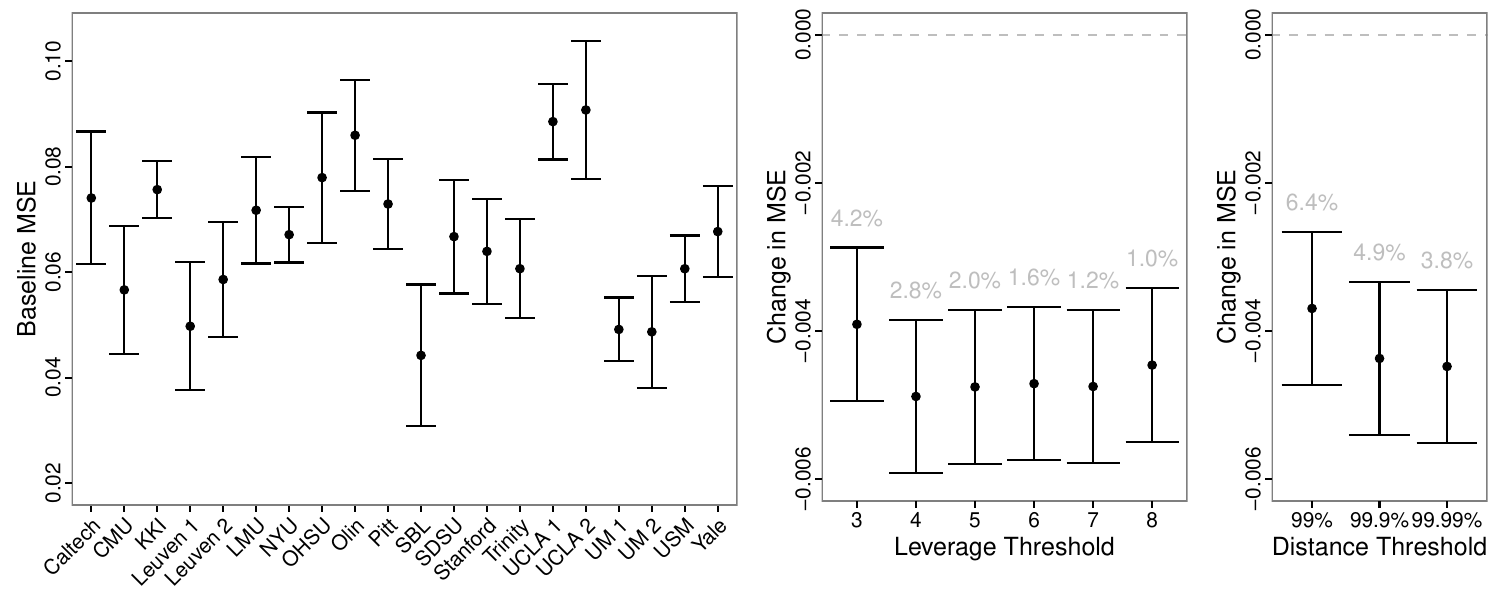}
\caption{Reliability of Between-Network Connectivity}
\end{subfigure}
\caption{Estimates and 95\% confidence intervals for the model coefficients for (a) the scan-rescan overlap of brain networks and (b) the scan-rescan MSE of connectivity between each network.  For both models, the left-hand plot dislays the fixed effects for each dataset ($\gamma_k$) and illustrates the heterogeneity in reliability across datasets in ABIDE before outlier detection. The middle and right-hand plots display the coefficients for each outlier removal method ($\alpha_m$), which represent the change in reliability due to outlier removal.  These plots also show the percentage of volumes in each fMRI run labeled as outliers using each method.  Both leverage and robust distance-based outlier removal methods result in a statistically significant improvement to reliability of brain networks and connectivity.  While both methods appear to be fairly robust to the choice of threshold, reliability is maximized by choosing a cutoff of $4$ times the median for PCA leverage and the $99.99$th quantile for PCA robust distance. \\[10pt]}
\label{overlap_coef}
\end{figure}

Figure \ref{overlap_coef} displays estimates and 95\% confidence intervals for the coefficients of the models for reliability of resting state networks (a) and functional connectivity (b).  For (a), larger overlap values represent greater reliability; for (b), smaller MSE values represent greater reliability.  The left-hand panels of (a) and (b) display the fixed effects for each dataset ($\gamma_k$) and illustrate the heterogeneity in baseline reliability across ABIDE datasets before outlier removal.  This reflects the substantial differences in acquisition, processing and quality control methods across the data collection sites contributing to ABIDE.  The middle and right-hand panels of (a) and (b) display the coefficients for each outlier removal method ($\alpha_m$).  The average percentage of volumes labeled as outliers using each method is also displayed in gray.  Both leverage-based and distance-based methods significantly improve reliability of estimates of subject-level brain networks and functional connectivity.  Improvement in reliability is maximized using a threhold of $\alpha=4$ or $5$ for leverage-based outlier removal and $1-\gamma=99.99$ for distance-based outlier removal.  The maximum improvement in overlap of resting-state networks is approximately $100$ voxels using either method, while the maximum reduction in MSE of functional connectivity is $0.005$ and is achieved using leverage-based outlier removal with $\alpha=4$. 

We also stratify the model by those subjects who passed quality inspection and those who failed.  Figure 1 of Supplementary Materials Appendix C shows estimates and 95\% confidence intervals for the model coefficients after stratification.  While subjects who failed quality inspection tend to improve more than those who passed quality inspection, differences are not statistically significant, and both groups of subjects benefit substantially from outlier removal.   

\section{Discussion}

We have proposed a method to detect outlying time points in an fMRI scan by drawing on the traditional statistical ideas of PCA, leverage, and outlier detection.  The proposed methods have been validated through simulated data and a large, diverse fMRI dataset.  We have demonstrated that the proposed methods are accurate and result in improved reliability of two common types of analysis for resting-state fMRI data, namely identification of resting-state networks through ICA and estimation of functional connectivity between these networks.

The proposed techniques are, to the best of our knowledge, the first to provide a single measure of outlyingness for time points in an fMRI scan, which can be easily thresholded to identify outliers.  Unlike motion-based outlier detection methods for fMRI, they are agnostic to the source of artifact and therefore may be used as a general method to detect artifacts, including those unrelated to motion.  Furthermore, PCA leverage is directly related to the estimation of principal components, which are an important quantity in the analysis of resting-state fMRI, as they are used as input to ICA for the identification of resting-state networks. 

One limitation of our approach is that we perform validation on a single dataset, the ABIDE.  However, this dataset is in fact a diverse collection of 20 datasets from 16 international sites, which strengthens the generalizability of our results.  Another limitation of the proposed methods is that they may be sensitive to the number of principal components retained.  However, we have found that the method performs well with different model orders (e.g. $20$ or $30$), and we propose an automated method of selecting model order in order to provide a fully automated approach.  

A limitation of PCA leverage-based outlier removal is that the proposed thresholding rule is, as in regression, somewhat ad-hoc.  However, the use of the median leverage across observations as a benchmark is a reasonable approach, and we have tested a range of threholds.  Based on our reliability analysis we expect a threshold of $4$ of $5$ times the median leverage to work well in practice for fMRI, but for different types of data the researcher may wish to re-evaluate this choice.  In particular, fMRI volumes containing artifacts tend to be very different from those volumes free of outliers, so a relatively high threshold for leverage and robust distance tends to work well; in other contexts, outliers may be more subtle.

While the proposed methods have been designed and validated for resting-state fMRI data, they may be easily extended to other types of medical imaging data, such as task fMRI and EEG data, as well as other types of HD data.  Furthermore, they may also be extended to group analyses. Future work should focus on exploring these directions.

As the availability of large fMRI datasets continues to grow, automated outlier detection methods are becoming essential for the effective use of such data.  In particular, the reliability of analyses employing these diverse datasets may be negatively impacted by the presence of poor quality data.  The outlier detection methods we propose have the potential to improve the quality of such datasets, thus enhancing the possibilities to use these data to understand neurological diseases and brain function in general.

\section*{Supplementary Materials} The reader is referred to the on-line Supplementary Materials for additional technical details.  Appendix A contains details regarding the fMRI data acquisition, pre-processing, and quality control procedures.  It also contains information about each dataset forming the ABIDE.  Appendix B details the estimation of subject-level resting-state networks through ICA and dual regression.  Appendix C displays the results of reliability analysis after stratifying by subjects whose scans passed quality inspection and those whose did not.

\section*{Funding} This work was supported by the National Institutes of Health [R01 EB016061], the National Institute of Biomedical Imaging and Bioengineering [P41 EB015909], and the National Institute of Mental Health [R01 MH095836]. {\it Conflict of Interest}: None declared.

\bibliographystyle{biorefs}
\bibliography{mybib}

\newpage

\end{document}